\documentclass[]{spie}  
\usepackage{graphicx}
\usepackage{dcolumn}
\usepackage{color}

\usepackage{amsmath}
\usepackage{amssymb}

\usepackage[tight,normalsize,sf,SF]{subfigure}
\title{Silicon-Based Antenna-Coupled Polarization-Sensitive Millimeter-Wave Bolometer Arrays for Cosmic Microwave Background Instruments}

\author{Karwan Rostem\supit{a,b}, Aamir Ali\supit{a}, John W. Appel\supit{a}, Charles L. Bennett\supit{a}, Ari Brown\supit{b}, Meng-Ping Chang\supit{c}, David T. Chuss\supit{d}, Felipe A. Colazo\supit{b}, Nick Costen\supit{c}, Kevin L. Denis\supit{b}, Tom Essinger-Hileman\supit{a}, Ron Hu\supit{c},  Tobias A. Marriage\supit{a}, Samuel H. Moseley\supit{b}, Thomas R. Stevenson\supit{b}, Kongpop U-Yen\supit{b}, Edward J. Wollack\supit{b}, Zhilei Xu\supit{a}
	\skiplinehalf
	\supit{a}Johns Hopkins University, Department of Physics and Astronomy, 3400 North Charles Street, Baltimore, MD 21218; \\
	\supit{b}Goddard Space Flight Center, 8800 Greenbelt Road, MD 20771; \\
	\supit{c}SGT Stinger Ghaffarian Technologies, 7701 Greenbelt Rd, MD 20770; \\
	\supit{d}Villanova University, Department of Physics,  800 E. Lancaster Avenue, Villanova, PA 19085. \\
}

\authorinfo{Further author information: (Send correspondence to K. Rostem)\\ E-mail: karwan.rostem@nasa.gov, Telephone: +1 301 286 0308}  

\begin{document} 
\maketitle 
 
\begin{abstract}
We describe feedhorn-coupled polarization-sensitive detector arrays that utilize monocrystalline silicon as the dielectric substrate material. Monocrystalline silicon has a low-loss tangent and repeatable dielectric constant, characteristics that are critical for realizing efficient and uniform superconducting microwave circuits. An additional advantage of this material is its low specific heat. In a detector pixel, two Transition-Edge Sensor (TES) bolometers are antenna-coupled to in-band radiation via a symmetric planar orthomode transducer (OMT). Each orthogonal linear polarization is coupled to a separate superconducting microstrip transmission line circuit. On-chip filtering is employed to both reject out-of-band radiation from the upper band edge to the gap frequency of the niobium superconductor, and to flexibly define the bandwidth for each TES to meet the requirements of the application. The microwave circuit is compatible with multi-chroic operation. Metalized silicon platelets are used to define the backshort for the waveguide probes. This micro-machined structure is also used to mitigate the coupling of out-of-band radiation to the microwave circuit. At 40 GHz, the detectors have a measured efficiency of $\sim$90\%. In this paper, we describe the development of the 90 GHz detector arrays that will be demonstrated using the Cosmology Large Angular Scale Surveyor (CLASS) ground-based telescope.

\end{abstract}


\section{\label{sec:intro}INTRODUCTION}

The polarization of the Cosmic Microwave Background (CMB) holds important clues about the primordial history of the universe. Within the concordance model, the inflationary scenario predicts that primordial gravitational waves (tensor modes) will source polarization in the CMB.  Moreover, these tensor modes will uniquely produce a divergence-free polarization pattern, known as ÒB-modesÓ that if detected and characterized, would have profound implications for cosmology and particle physics at energy scales near that predicted for grand unification.

At angular scales below 2$^\circ$, the B-mode polarization is mainly sourced by gravitational lensing of the CMB from large scale structure. At large scales, the B-mode polarization is expected to be dominated by the inflationary signal, and is anticipated to be $\sim$1 nK above the 2.725 K isotropic CMB. Thus, large arrays of efficient background-limited detectors are needed for high-fidelity polarization measurements of the CMB. An additional challenge is the presence of foregrounds. Galactic synchrotron and dust emission produce polarized signals at millimeter wavelengths that are large compared to that of the CMB. The synchrotron component dominates at low frequencies, and the dust emission at high frequencies. To separate these, it is essential to pursue multi-band observations. Thus, the detectors utilized must be implemented across multiple bands. 

Given the low signal-to-noise ratio and large foreground contribution, the detectors are required to have low systematic effects arising from the optical coupling scheme (cross-polarization) and electro-thermal response (differential gain). The optical mode coupling to the bolometers is an essential element of the response that if not controlled could lead to low in-band efficiency, and sensitivity to out-of-band (stray light) signal. The bolometers should also be stable to allow polarization modulation of order a few Hz. 

Numerous ground-based and balloon-borne missions are under way to measure the large angular scale spectrum of the CMB polarization~\cite{SPIDER2010,GroundBIRD2012,PIPER2014}. In addition, several space instruments have been proposed to realize the ultimate promise of CMB polarization science, most recently LiteBIRD~\cite{LiteBIRD} and COrE+~\cite{COrEPlus}. The large format detector arrays needed for these missions must be compatible with the space environment. Semiconducting bolometers have been employed in space in the High Frequency Instrument of the Planck mission~\cite{Holmes2008}. The detectors registered a higher than expected rate of events induced by the absorption of cosmic ray (ionizing) particles at the focal plane~\cite{Catalano2014,Catalano2014v2,Miniussi2014}. This experience is valuable for future space-based mm-wave instrumentation, and suggests the need for mitigating strategies such as ensuring high bulk electrical and thermal conductivity where possible, and minimizing the footprint of high-quality dielectric materials to avoid the negative impact of surface and deep dielectric charging. 

In this paper, we report on the progress of feedhorn-coupled silicon-based polarization-sensitive detectors for 90 GHz observations. The microwave circuit has a flexible design that can be tailored for single frequency and multichroic detection up to 300 GHz. Targeting operation in a space instrument, the array design is integrated with several features that mitigate the negative impact of dielectric charging. 

We are demonstrating the silicon-based detector technology on the ground-based instrument the Cosmology Large Angular Scale Surveyor~\cite{Essinger2014,Appel2014,rostem:SPIE2014}. CLASS has four receivers operating at 40, 90, 150, and 220 GHz. The detectors for the two highest frequency channels will be implemented with a dichroic planar circuit. Previously, we reported on the fabrication~\cite{denis:fab} and testing of the single pixel 40 GHz detectors~\cite{Appel2014,rostem:precision}. CLASS has started observations with the 40 GHz focal plane receiver. Preliminary results indicate the optical loading is within the margins estimated from the atmosphere and instrument emission~\cite{Essinger2014}. CLASS, which is located at the Atacama Desert in Chile, observes $\sim$70\% of the sky every day and recover the polarization power spectrum at angular scales above 2$^\circ$. 

In Section~\ref{sec:90GHz}, we describe the 90 GHz detector circuit and array architecture that includes a modular integrated backshort and stray light control enclosure. In Section~\ref{sec:space}, we describe the properties of the array that mitigate the effects of dielectric charging in a space environment.  In Section~\ref{sec:crossover}, we provide a survey of microwave crossover technology and describe an airbridge scheme that we have developed for broadband operation, from DC-500 GHz. In Section~\ref{sec:packaging}, we describe a packaging scheme developed for the CLASS receiver, and report on the preliminary optical properties of a 90 GHz array. We refer to the paper by Harrington et al.~\cite{Harrington2016} in these proceedings for a detailed discussion of the CLASS science goals and observatory status. 

\section{90 GHz Detector Array}
\label{sec:90GHz}

\begin{figure}[!t]
\begin{center}
\subfigure[]{\includegraphics[height=7cm]{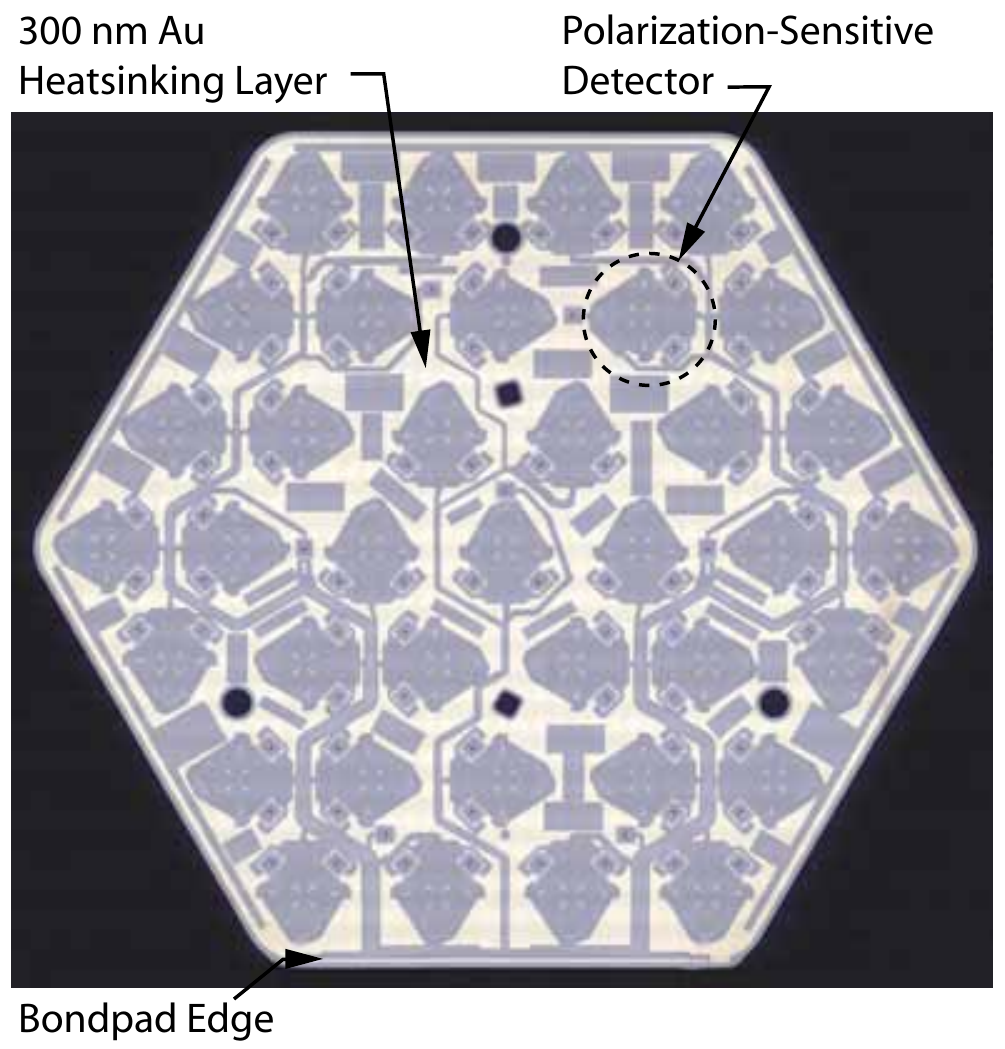}}
\subfigure[]{\includegraphics[width=8cm]{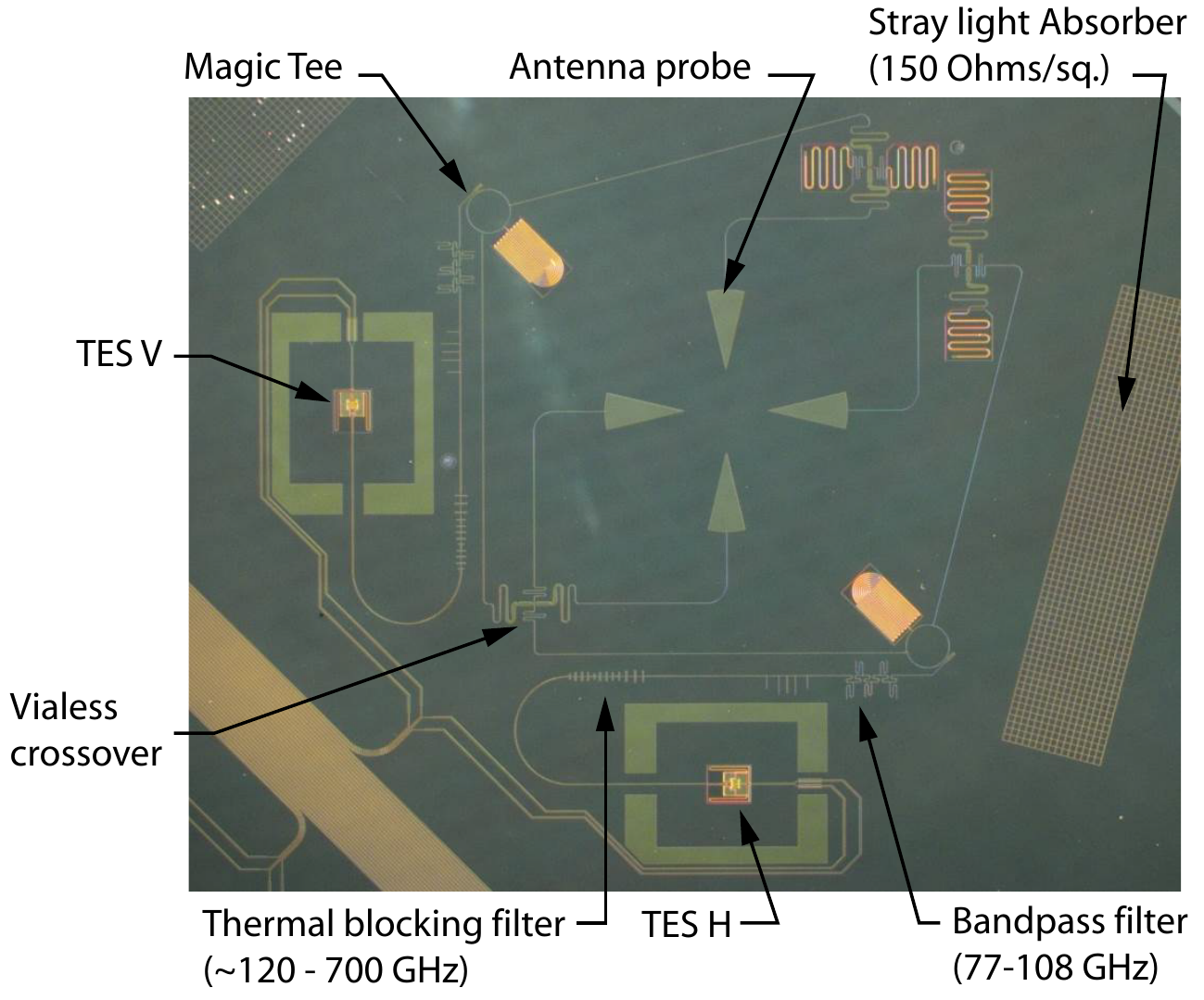}}
\caption{\label{fig:90GHzModule}(a)~Photograph of a silicon-based 90 GHz detector array with 37 dual-polarization sensitive detectors, circuit side. (b)~Zoomed image of the detector circuit showing the various microwave components described in the text. }
\end{center}
\end{figure}

The feedhorn coupled planar detector circuit has been described previously in the context of the 40 GHz detector development for CLASS~\cite{rostem:precision,Essinger2014,rostem:SPIE2014}. Here we briefly review the fabrication and circuit schematic. In contrast to the lower frequency channel, the 90 GHz detectors are fabricated on 7 hexagonal modular arrays. Each array is patterned out of a 100 mm silicon-on-insulator wafer and holds 37 dual-polarization sensitive detectors as shown in Fig.~\ref{fig:90GHzModule}(a). A 90 GHz modular array consists of three wafers: the silicon-on-insulator detector wafer with a 5 $\mu$m thick float-zone single-crystal silicon layer, a spacer wafer that is approximately quarter-wave thick in-band, and a cap wafer that forms a short for the orthomode transducer (OMT) termination. The single-crystal silicon device layer was chosen for its excellent microwave~\cite{datta:siliconloss} (low loss and repeatable dielectric function), mechanical (stress free), and thermal~\cite{rostem:precision} (negligible heat capacity at 100 mK and below) properties. 

As shown in Fig.~\ref{fig:90GHzModule}(b), a planar OMT circuit is used to couple light to two Transition-Edge Sensor bolometers. The OMT probes couple to the TE$^\bigcirc_{11}$ mode of a circular waveguide. Signal from opposing probes are differenced in a 180$^\circ$ hybrid~\cite{yen:magicTee} that terminates the common mode signal in a Au broadband resistor termination. The difference signal is capacitively coupled to a microstrip line that is filtered for bandpass definition, out-of-band signal rejection, and eventually terminated at a PdAu resistor on the TES bolometer membrane~\cite{rostem:SPIE2014}. The microwave circuitry and TES bias leads are fabricated out of superconducting niobium. At the operating bath temperature of 0.1 K, there is negligible microwave loss and Joule heating. Normal metal components such as Au and PdAu are used to terminate microwave signals, and to effectively define the TES detector time constant through the electronic heat capacity of Pd~\cite{rostem:SPIE2014}. 

\begin{figure}[!t]
\begin{center}
{\includegraphics[width=17cm]{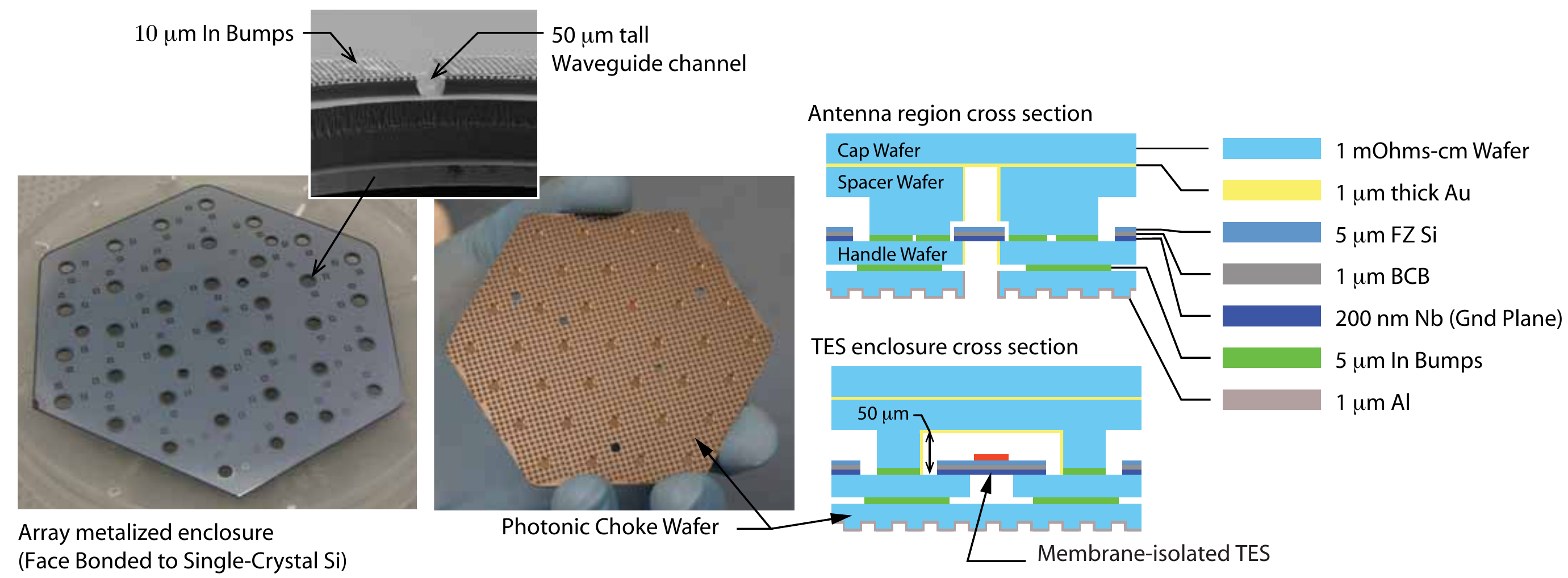}}
\caption{\label{fig:Backshort}~(Left) Photograph of a metalized enclosure that forms the backshort termination for the 37 orthomode transducers and suppresses the coupling of stray light to the microwave circuit and the TES bolometers. (Center) Photonic choke~\cite{yen:pcj} wafer. The surface and waveguide holes are coated with 1 $\mu$m of Al. (Right) Cross section of the whole array when bonded together. The high quality dielectric is 5 $\mu$m of float-zone single-crystal silicon. }
\end{center}
\end{figure}

In a separate fabrication chain, the spacer and cap wafers are processed and bonded together to form a single modular enclosure (see Fig.~\ref{fig:Backshort}). For the spacer, circular waveguide holes and relief structures in the form of 50 $\mu$m tall waveguide channels are formed using deep reactive ion etching on the side facing the single-crystal silicon layer. Indium bump bonds are also realized with a liftoff process around each channel. The sidewall of the circular holes and the surface of the cap wafer are coated with Au to ensure good conduction. The spacer and cap wafer are then bonded together using Au-Au thermo-compression bonding. These steps along with the fabrication of the detector wafer and circuitry is described in detail elsewhere~\cite{Denis2015}.

The single-crystal silicon layer is the only high-resistivity dielectric in an array. The rest of the silicon layers are degenerately-doped silicon (DDS) with $< 3$ mOhm-cm resistivity at room temperature. This choice reflects the practical aspect of the fabrication. DDS ensures that the electrical volume is confined to the microwave transmission lines. Thus, it prevents the formation of unwanted microwave cavities when fabrication imperfections in the metal coatings occur at the waveguide features of the modular enclosure. As a result, the microwave circuit is tolerant to changes in the metal coatings that may occur within an individual detector array, or across several arrays at the focal plane of an instrument.

On the detector wafer, the high-resistivity silicon, Nb ground plane, and a BCB bonding layer are etched to expose the low-resistivity handle wafer. The hybridized cap-spacer wafer is then bonded to the handle wafer using flip chip indium bonding~\cite{Denis2015}. The final structure incorporates the approximate quarter-wave backshort, and forms both the boxed microstrip waveguide where the gold coated channels are etched and a metalized housing for the TES detectors as illustrated in Fig.~\ref{fig:Backshort}.

\section{Design considerations for space}
\label{sec:space}

\subsection{Dielectric Charging}

The space environment presents distinct challenges to the operation of an instrument. The internal and external surfaces of a spacecraft can be subject to a charge imbalance depending on the orbit, plasma temperature, solar and geomagnetic activity, sunlight/dawn exposure, and shielding~\cite{Mizera1983,Leung1986,Frederickson1996}. Unintentional charge accumulation can potentially have detrimental effects over time. If the voltage difference across a dielectric or between adjacent materials exceeds the breakdown voltage, an electrostatic discharge may occur. These events may readily induce electromagnetic interference that can upset spacecraft operation~\cite{Leung1982}. Without appropriate charge control, discharge or arcing can potentially damage the insulating materials. 

To mitigate charge buildup and the effects of electrostatic discharge events on the detector array, the principal strategy should be to minimize the occurrence of floating conductors, and the footprint of high quality dielectric substrate materials needed for the cryogenic mm-wave sensors. Charge control measures should then be implemented to electromagnetically shield the remaining circuit elements. 

In the detectors described here, the single-crystal silicon is utilized simultaneously for structural membranes and as a microwave substrate. It is the only high-quality dielectric in the detectors, and its exposure to the environment is limited. The DDS layers are in electrical contact with the circuit ground plane through a via etched in the single-crystal silicon, Nb ground plane, and BCB bonding layer, as described in Sec.~\ref{sec:90GHz} and illustrated in Fig.~\ref{fig:Backshort}. To ensure the detector signal quality is not corrupted by electromagnetic interference from discharge events~\cite{Leung1982}, the ground plane is accessible from the front side of the detector circuit through a via hole, and can be wirebonded to a readout circuit ground to provide a common ground for the DC and microwave circuits. In addition to these measures, the DDS layers form a stray light control enclosure that limits the coupling of out-of-band radiation to each TES bolometer, as illustrated in Fig.~\ref{fig:Backshort}. A waveguide channel with conductive walls is formed when the backshort wafer is bonded to the detector handle wafer. Electrical access to the TES enclosure is thus through a boxed microstrip line. 

The signal conditioning extends from RF up to the energy gap frequency of the niobium microwave circuitry at $\sim$700 GHz. Above this frequency, the resistive properties of the planar Nb transmission lines acts as a lossy filter. To provide further heat and charge transfer paths across the single-crystal silicon layer, a Au layer is deposited on the detector side as shown in Fig.~\ref{fig:90GHzModule}(a). 

\subsection{Ionizing Radiation}

Particles with energy ($> 10$ keV) can penetrate the bulk of an insulating material and lead to deep dielectric charging~\cite{Leung1986}. If the discharge rate is less than the deposition rate, voltage differences can reach breakdown limits. Cosmic rays and and high energy protons released during solar activity form the majority of the particles that are responsible for deep dielectric charging. The particles lose energy through ionization as they travel through an insulating layer, releasing electron-hole pairs and phonons. The charge buildup could be managed by the same strategies outlined earlier in the context of surface charging. In addition, dissipative elements and resistive coatings can be employed to tailor the charge dissipation rate independently of the microwave circuit. 

Of equal concern to a bolometer are the phonons released during an ionizing event. Depending on the location of the event with respect to the bolometer, material surface asperity, material interface, presence of conductive coatings, and geometry, a fraction of the phonons may travel ballistically and could reach the bolometer, which senses the phonon flux as an optical signal. The bolometers on the HFI instrument of the Planck mission were reportedly affected by a higher than expected rate of ionizing particle events, leading to significant data loss ($\sim$20\%). Detailed studies~\cite{Miniussi2014,Catalano2014,Catalano2014v2} have shown that the events could be characterized by the location of the absorption, such as the semi-conducting thermistor, absorber grid, and silicon die to name a few. To mitigate the effect of ionizing particles, in addition to minimizing the footprint of high-quality dielectrics, we implement structures having simple (easily understood) and fast thermal response time. The Au coating on the single-crystal silicon layer visible in Fig.~\ref{fig:Backshort} increases the probability of thermalizing ballistic phonons before they reach the TES bolometers.

Although the Planck semi-conducting bolometers had a detector time constant of $\sim$5 ms, stable background-limited TESs could be designed with a factor of five improvement in speed. This is another strategy that can be adopted to significantly reduce the affected portion of the data from a bolometer in a space instrument. 


\section{Broadband Microwave Crossover}
\label{sec:crossover}

The planar OMT employs a microwave signal crossover to route the orthogonal polarization signals. Broadly speaking, a crossover can be realized using a physical connection such as a bridge or thru via~\cite{Taylor1986,Horng1994,Pang2016}, or a vialess geometry in which mode conversion~\cite{Lin2001,Yen2009,Abbosh2012} and symmetry~\cite{deRonde1982,Chen2007} enable the needed port coupling and isolation. The latter technique can be further classified into two subcategories, in which either delay lines in the form of a 4-port rat race~\cite{deRonde1982,Chen2007} or microstrip-to-slotline mode conversion are used~\cite{Lin2001,Yen2009,Abbosh2012}.  

\begin{figure}[!b]
\begin{center}
{\includegraphics[width=17cm]{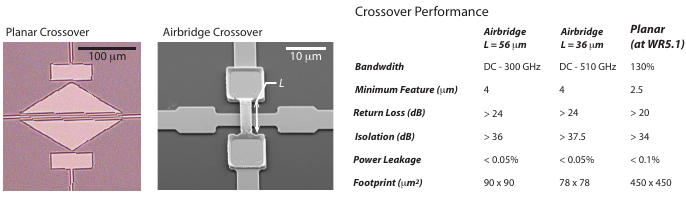}}
\caption{\label{fig:Crossover}~(Left) Photograph of a broadband vialess crossover based on the design of Abbosh et al.~\cite{Abbosh2012}. (Middle) A niobium air bridge crossover developed for DC-500 GHz operation. (Right) Table  comparing the microwave and geometric properties of the two crossovers. The waveguide standard WR5.1 band is 140-220 GHz.}
\end{center}
\end{figure}

We have implemented a microstrip-to-slotline vialess crossover at 40 and 90 GHz over a 2:1 waveguide bandwidth~\cite{Yen2009}. Many factors play a role in determining the coupling, radiative loss, and isolation of a 4-port microstrip-to-slotline vialess network. The loss could be reduced by minimizing the slot width to guide wavelength ratio, and the physical footprint of the crossover geometry. For broadband operation above 150 GHz, the required slot width approaches $\sim$2 $\mu$m. Figure~\ref{fig:Crossover} shows a broadband vialess crossover design~\cite{Abbosh2012} suitable for operation at 150 and 220 GHz. The advantage of this design is the improved isolation due to a tapered coplanar waveguide mode converter, and reduced radiative loss due to its more effectively lumped geometry. 

Figure~\ref{fig:Crossover} also includes the geometric and microwave properties of two crossover devices. For broadband operation and minimal radiative loss, an airbridge crossover is most effective, at the cost of increased fabrication steps, though not necessarily difficulty. Figure~\ref{fig:Crossover} shows an image of a niobium airbridge crossover suitable for DC-500 GHz operation.

The airbridge is fabricated with a standard polymer based sacrificial layer process. First, 200 nm of niobium is sputter deposited and reactive ion etched in a CF$_4$/O$_2$ plasma. A layer of photoresist (SPR-220) is then spun, and the via is patterned and developed. The photoresist serves as a sacrificial layer and sets the spacing between the niobium layers. A high temperature baking step above the glass transition temperature of the resist reflows the edges of the photoresist to improve metallization step coverage. Next, the 500 nm thick second layer of niobium is deposited after an in-situ reverse bias sputter clean to remove any residual photoresist and niobium oxide. During DC sputtering deposition, the target is reverse biased with an RF power supply in order to improve step coverage. The second layer of niobium is patterned with the same resist as the sacrificial layer and reactive ion etched with CF$_4$/O$_2$.  Finally, the resist is dissolved in acetone releasing the air-bridge crossover layer. Subsequent cleaning in high pressure low power oxygen plasma removes residual photoresist. 

\section{Packaging and Optical Response}
\label{sec:packaging}



\begin{figure}[!t]
\begin{center}
\subfigure[]{\includegraphics[height=7cm]{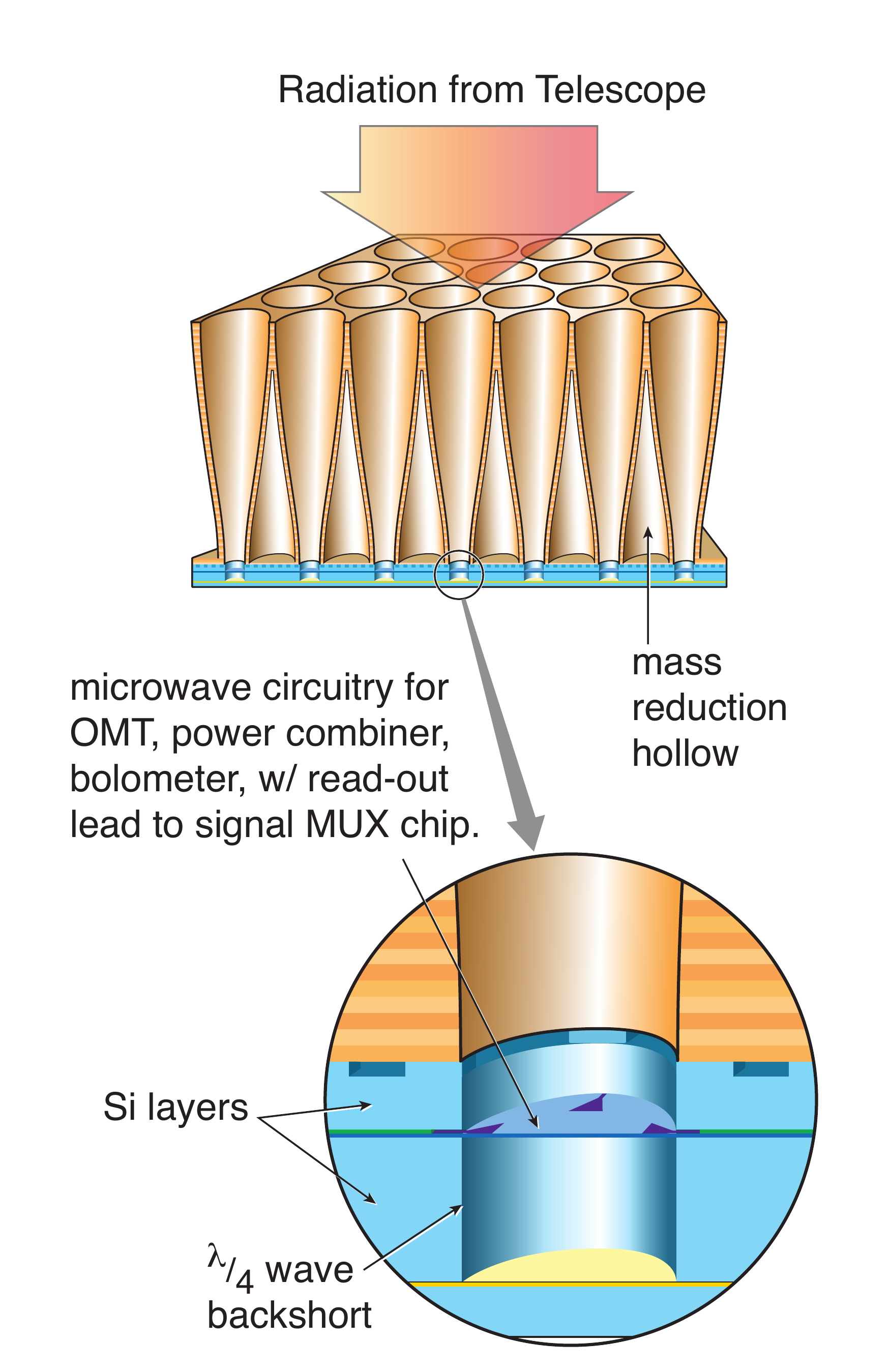}}
\hspace{10mm}
\subfigure[]{\includegraphics[height=7cm]{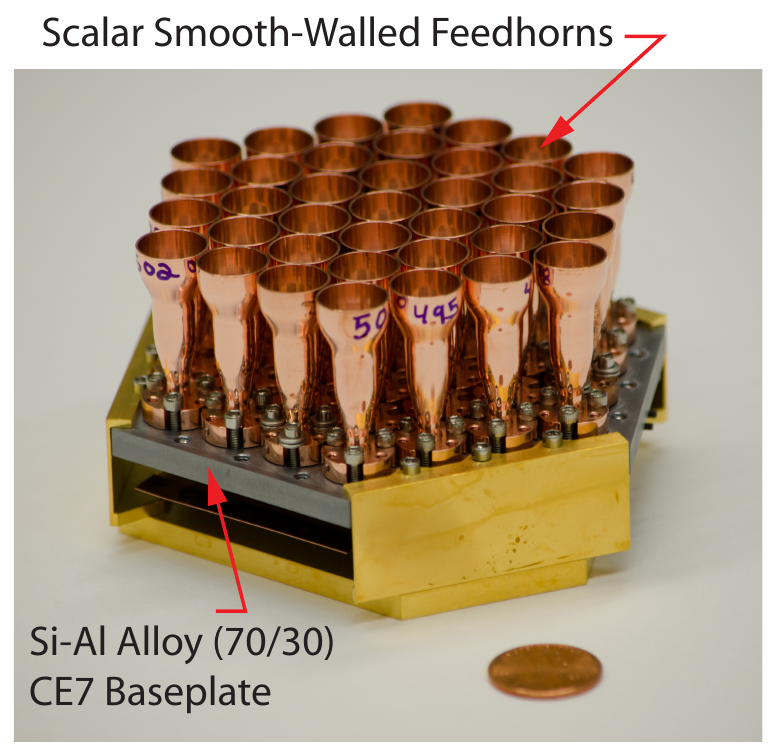}}
\caption{\label{fig:90GHzPackage}(a)~Simplified illustration of an array of feedhorn-coupled detectors. The photonic choke pillars are part of a silicon detector array. (b)~A 90 GHz modular package assembly. The detector and readout circuitry are vertically stacked to enable the compact format.  }
\end{center}
\end{figure}

The detector array shown in Fig.~\ref{fig:90GHzModule} can be housed in a hexagonal modular package. For a feedhorn-coupled scheme, the package includes an optical baseplate with precision waveguide features to couple the radiation to the OMT circuit at the detector plane. The optical baseplate and the silicon detectors must also be CTE matched, and the precision waveguide alignment needs to be maintained at cryogenic temperatures. Figure~\ref{fig:90GHzPackage}(a) illustrates the coupling scheme for a silicon detector array as shows in Fig.~\ref{fig:90GHzModule}~and \ref{fig:Backshort}. The photonic choke joint realized with the photonic choke wafer and the optical baseplate provides a virtual ground for the in-band radiation, and ensures a graceful degradation of the in-band leakage with increased spacing between the photonic choke wafer and the baseplate~\cite{yen:pcj}. Thus, the detector packaging forms an essential part of the electromagnetic circuit. 

Figure~\ref{fig:90GHzPackage}(b) shows an example of a modular package developed for the CLASS instrument. We have chosen an alloy of silicon (70\%) and aluminum (30\%) as the optical baseplate material. Generically referred to as CE7\footnote{Sandvik Osprey, controlled-expansion CE7 alloy.}, this alloy provides an excellent CTE match to the silicon detectors. CE7 is machinable and conductive at room temperature. Below 1.2 K, the material superconducts and becomes a poor thermal conductor. Though not shown in Fig.~\ref{fig:90GHzPackage}, to improve the thermal propagation, the CE7 is plated with gold. For a space instrument, a silicon platelet architecture can be implemented to realize a CTE-matched thermally conducting baseplate that also serves as a corrugated feed~\cite{OpticalCoupling2009,Britton2010}. The cryogenic multiplexing readout circuit for the TES bolometers is vertically stacked in the package housing.
 



We have optically characterized a 90 GHz array with a Fourier Transport Spectrometer. Preliminary results suggest the measured band edges are in good agreement with our models of the bandpass filter defining the in-band signal. We have found it necessary to model the superconducting Nb transmission lines as 3D objects in a full-wave simulator. To speed up the modeling of the full circuit shown in Fig.~\ref{fig:90GHzModule}(b), first we model a simple microstrip transmission line with a 3D top conductor, where the inner volume is vacuum and a surface impedance calculated from BCS theory is applied to the surfaces of the microstrip. A correction term is applied to the surface impedance to account for the finite thickness of the conducting sheets representing the microstrip~\cite{Kerr1996}. In a separate model, the surface impedance of a zero-thickness top conductor microstrip is tuned by a constant factor until the characteristic impedance and phase velocity match the results from the 3D model across the simulation band. The BCS kinetic inductance is estimated from knowledge of the Nb thin-film residual-resistivity ratio and zero temperature London penetration depth. This type of circuit modeling has been shown to agree well with measurements of the microwave response. For a 90 GHz detector circuit, the band pass edges agree to within $\sim$2 GHz. Test are ongoing to measure the absolute value of the in-band signal efficiency using a cryogenic blackbody calibrator. 

\section{Conclusion}

We have described the development of feedhorn-coupled polarization-sensitive detector arrays suitable for Cosmic Microwave Background observations. The detectors are fabricated on single-crystal silicon, a low loss dielectric with repeatable electromagnetic, mechanical and thermal characteristics. The microwave circuit uses a planar orthomode transducer to couple light to two Transition-Edge Sensor bolometers. The detector array includes a modular conducting enclosure that limits the coupling of out-band stray light radiation to the microwave circuitry and the bolometers. The enclosure also serves to define the backshort termination for the orthomode transducers across the array. The comprehensive filtering strategy, stray light control, and metallic beam-forming elements (feedhorns) minimize the need for quasioptical dielectric elements in a space instrument design. 

The array architecture has been designed, from the outset, to mitigate the effect of surface and deep dielectric charging experienced in a space instrument. The single-crystal silicon dielectric is the only high-quality substrate material in an array. The metalized degenerately-doped silicon layers that form the modular enclosure are in contact with the microwave ground. Floating conductors are avoided throughout the design. The microwave ground reference can be defined and controlled with respect to the detector readout through the use of a via. We have also developed a superconducting niobium airbridge crossover suitable for DC-500 GHz operation. 

The detector arrays will be demonstrated in the ground-based receivers of the Cosmology Large Angular Scale Surveyor. CLASS has four receivers, a low frequency 40 GHz channel to remove the synchrotron foreground, two CMB receivers at 90 GHz, and a high-frequency dust channel that will simultaneously observe at 150 and 220 GHz. The detectors for the 40 GHz focal plane have been fabricated and saw first light in mid-May 2016. We have characterized the optical response of a 90 GHz pixel in one of these arrays using a Fourier Transform Spectrometer. The results show the microwave circuit response agrees well with a model of the full circuit. Tests are ongoing to determine the in-band signal efficiency using a cryogenic blackbody source. The detectors for the higher frequency channels are currently in fabrication and will be deployed over the next two years

\section{Acknowledgement}

The detector technology developments were supported by NASA/ROSES APRA funding. We acknowledge the National Science Foundation Division of Astronomical Sciences for their support of CLASS under Grant Numbers 0959349 and 1429236. We thank Marc Castro for his contributions to the crossover design efforts. 


\end{document}